# Combinatorial sputter deposition of CrMnFeCoNi high entropy alloy thin films on agitated particles

Florian Lourens[a], Alfred Ludwig[a,] *

[a] *Materials Discovery and Interfaces, Institute for Materials, Ruhr University Bochum, Universitätsstrasse 150, 44780 Bochum, Germany*

*\* Corresponding author: alfred.ludwig@rub.de*

**Abstract**

A method for combinatorial sputter deposition of thin films on microparticles is presented. The method is developed for a laboratory-scale magnetron sputter system and uses a piezoelectric actuator to agitate the microparticles through oscillation. Custom-made components enable to agitate up to nine separate batches of particles simultaneously. Due to the agitation, the whole surface of the particles can be exposed to the sputter flux and thus can be completely covered with a thin film. By sputtering a CrMnFeCoNi high entropy alloy target, separate batches of polystyrene microspheres (500 μm monodisperse diameter), Fe alloy particles (300 μm mean size) and NaCl salt particles (350 μm mean size) were simultaneously coated with a homogeneous thin film. In contrast, a CrMnFeCoNi thin film that was deposited on agglomerating Al particles (5 μm mean size) only partially covers the surface of the particles. By co-sputtering a CrMn, an FeCo and a Ni target, nine separate batches of Al particles (25 μm mean size) were coated with a CrMnFeCoNi thin film with a composition gradient. These depositions demonstrate the ability to coat different types of particles with uniform films (from elemental to multinary compositions) and to deposit films with composition gradients on uniform particles.

# 1. Introduction

Powders that consist of microparticles are used in various industries and current fields of research such as foods, pharmaceuticals, cosmetics, powder metallurgy, additive manufacturing, lacquers and catalysis [1–3]. Many powder properties such as flowability [3], electrocatalytic activity [4,5], compactibility [9] and oxidation resistance [6] depend completely or to a great extent on the surface properties of the particles. Since many materials with desirable surface properties are valuable, core-shell particles made of relatively cheap core particles and functional coatings are of high scientific and technological interest. Among different methods to produce core-shell particles, several studies demonstrate sputter deposition as an effective means to coat particles to alter the properties of powders. For example, stainless steel coatings on WC particles used in additive manufacturing improve the flowability and the oxidation resistance of the WC powder [3,6]. Ni thin films deposited on quasi-crystalline Al-Cu-Fe microparticles act as binder to enable cold-pressing the powder into dense macrocomposite cylinders, which is otherwise accompanied by crumbling or cleavage [9]. Several studies demonstrate the fabrication of cost-efficient catalytic powders by depositing catalytic thin films on microparticles, e.g. Pt coatings on polymer and $Al_2O_3$ particles, which show the same electrocatalytic properties as bulk Pt [4,5].

To modify powder properties effectively, it is important that the coatings cover the whole surface of the individual particles. To achieve this with sputtering, the particles must be agitated during the deposition process. Due to the line-of-sight characteristic of sputtering, not-agitated particles are mainly coated on the exposed top surfaces whereas the bottoms are shadowed and remain uncoated. Prior literature comprises different agitation techniques that can be classified according to their basic principles into four groups: rotating cylinders [1–6], rotating tilted bowls [7,8], vibrating bowls [10–14] and dusty plasmas [9,15]. These agitation techniques are schematically depicted in Fig. 1. They were successfully used to homogeneously coat particles (2 … 500 μm size) with metallic and metal oxide thin films. In some cases, difficulties with stuck particles made it necessary to modify the technique. For example, Abe et al. modified a rotating cylinder to a hexagonal barrel, which improved the agitation of the $Al_2O_3$ particles (20 µm size) [1,4]. In [7], a concussion mechanism was added to a rotating tilted bowl to loosen stuck glass microspheres. The technique from Fernandes et al. is based on a rotating cylinder that simultaneously vibrates [16].

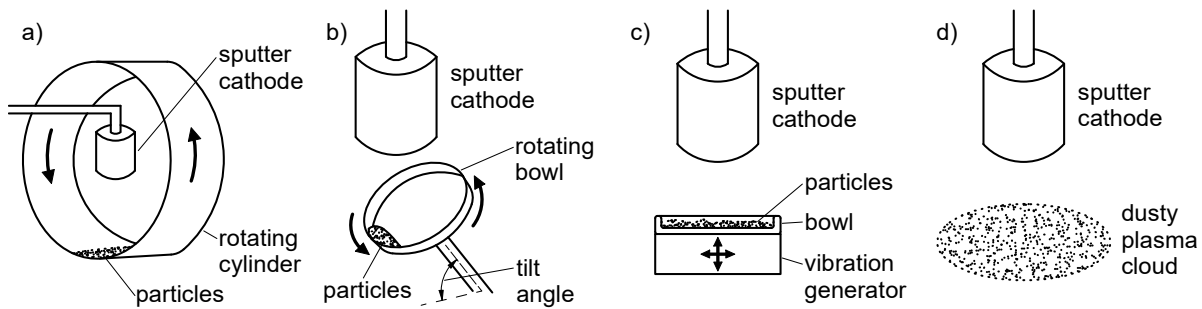

**Fig. 1:** Schematic illustrations of methods for sputter coating of particles, utilizing four particle agitation techniques: (a) rotating cylinder [1–6], (b) rotating tilted bowl [7,8], (c) vibrating bowl [10–14], (d) dusty plasma [9,15].

We introduce a method for combinatorial sputter deposition of thin films on agitated particles. This allows to deposit a homogeneous thin film in one process on different kinds of particles. Alternatively, a composition gradient film can be deposited on uniform particles via co-sputtering. With this approach, different combinations of particles and coating materials can be fabricated rapidly to accelerate research [17].

## 2. Development of the sputter method

Inspired by the agitation technique depicted in Fig. 1c, a device referred to as oscillation plate was developed. Fig. 2a shows a CAD model and Figs. 2b and 2c show photographs of the oscillation plate installed in a sputter chamber (AJA International, USA, 25 cm inner diameter). The oscillation plate consists of two custom-made stainless-steel components at the bottom, a piezoelectric actuator and two custom-made aluminum (Al) alloy components at the top. The two bottom components (56 mm diameter) are used for a fixed support of the oscillation plate to ensure a consistent position. The piezoelectric actuator (Physik Instrumente P-840.10V) comprises a stack of lead zirconate titanate (PZT) components that are spring-loaded inside a vacuum-compatible cylindrical housing. The head joint of the actuator is connected to the PZT stack and protrudes from the housing. Two components (50 mm diameter), which are custom-made of Al alloy for minimal weight, are bolted onto the head joint of the actuator. By applying an AC electrical signal, the PZT stack mechanically oscillates uniaxially in vertical direction with a frequency according to the driving signal. The driving signal is generated by a function generator (DS335, Stanford Research Systems) outside the vacuum chamber and fed in via a D-Sub electrical feedthrough. The exposed surface of the upper Al alloy component features nine milled pits (5 x 5 $mm^2$ horizontal ground area each, 1 mm depth) to carry up to nine separate batches of particles. Due to the fixed support of the oscillation plate, the oscillation generated from the actuator is directly transferred to the Al alloy components to agitate the loaded particles. The particles are positioned about 110 mm beneath three magnetron sputter cathodes (3.81 mm target diameter). Sputtering from one cathode results in a thin film with constant composition on the particles. However, a thickness gradient of the deposited thin film must be considered, because the sputter cathodes are not centered above the oscillation plate and the latter cannot be rotated due to its cabling. Co-sputtering of different target materials from two or three cathodes results in a thin film with a composition gradient across the particles.

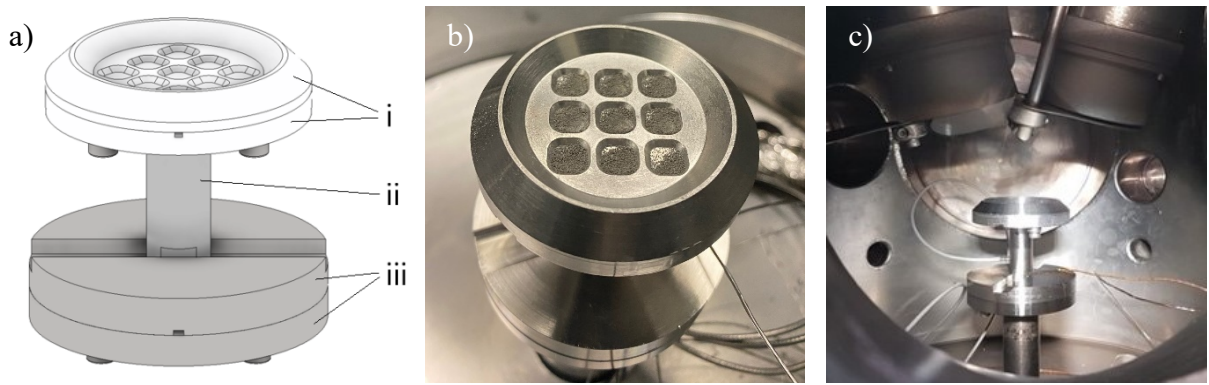

**Fig. 2:** (a) CAD model of the oscillation plate consisting of (i) two Al alloy components with nine milled pits for the substrate particles, (ii) a piezoelectric actuator and (iii) two stainless-steel components. (b, c) Photographs of the oscillation plate installed in the sputter chamber.

## 3. Experimental procedure

To test the developed method, the oscillation plate was used to agitate different types of microparticles in three deposition processes. The high entropy alloy CrMnFeCoNi was chosen as coating material, as it shows interesting structural [18] and electrocatalytic [19] properties. It has been investigated in thin film form on flat substrates [20,21] and also on nanotip arrays to study its phase stability [22,23]. The coated particles were investigated with scanning electron microscopy (SEM, Jeol JSM-7200F). The reported average chemical compositions (in at. %) of the deposited films refer to the average film composition of ten coated particles of the same sample and were measured by energy dispersive X-ray spectroscopy (EDX).

During the first deposition process, a CrMnFeCoNi high entropy alloy target was used and a thin film was deposited on Al particles (5 µm mean size). The sputter target was manufactured at Ruhr University Bochum and its chemical composition was determined by EDX as $Cr_{20.6}Mn_{20.4}Fe_{20.1}Co_{19.8}Ni_{19.1}$. An AC electrical signal with a square waveform and a peak-to-peak amplitude $V_{P-P}$ of 20 V was applied on the actuator. 20 V is the maximum $V_{P-P}$ of the function generator and was chosen because it caused the most visible movement in the particles. The frequency of the driving signal was set to a linear frequency sweep in the range of 0.1 … 4.0 kHz with a sweep rate of 1.0 Hz. It was observed that the frequency sweep caused more movement in the particles than a steady frequency. The sputter duration was 90 min and the sputter power was 30 W DC. The Ar (99.9999 % purity, Praxair) pressure during all three deposition processes was 0.5 Pa.

During the second deposition process, the same CrMnFeCoNi high entropy alloy target as in the first process was used and a thin film was deposited on three different powder substrates: polystyrene microspheres (500 µm monodisperse diameter), Fe alloy particles (300 µm mean size) and NaCl salt particles (350 µm mean size), which were placed in separate pits of the oscillation plate. The driving signal applied on the actuator was the same as in the first process, except for $V_{P-P}$, which was reduced to 7 V. At $V_{P-P} > 7$ V, the larger travel range of the actuator caused the particles to fall off the oscillation plate, whereas at $V_{P-P} < 7$ V, they moved visibly less. The sputter duration was 50 min and the sputter power was 30 W DC.

The third process was a co-deposition of $Cr_{50}Mn_{50}$ (99.8 % purity, EvoChem), $Fe_{50}Co_{50}$ (99.95 % purity, Robeko) and Ni (99.99 % purity, EvoChem). Nine batches of Al particles (25 µm mean size) were used as substrates (Fig. 2b). For a clear identification of the particle batches, they are numbered with respect to the sputter set up as depicted in Fig. 9a. The driving signal applied on the actuator was the same as in the first deposition process. The sputter duration was 45 min. The sputter powers were 25 W DC (CrMn), 30 W DC (FeCo) and 23 W DC (Ni).

## 4. Results and discussion

The first deposition process resulted in an uneven distribution of the sputtered elements on the Al particles (5 µm mean size), as shown in Fig. 3. This can be explained by the agglomerating characteristic of these Al particles. When the particles were agitated at ambient conditions, there was little movement in the agglomerates. At the sputter pressure, the movement was reduced and after a few minutes of sputtering, the agglomerates stuck to the oscillation plate and did not move at all. Consequently, only the particle surfaces exposed to the sputter flux got coated. A similar observation was reported by Schmid et al. [7]. In their case, glass microspheres (2 … 30 µm diameter) trickled freely in a rotating tilted bowl at ambient conditions, but stuck to the bowl at the sputter pressure of 1.5 Pa. The authors suggest that at ambient conditions, the air humidity leads to thin water layers that lubricate the particles to move freely, whereas at sputter conditions there are no such layers. They also report that the deposited metal and metal oxide coatings further promote the sticking. Sticking was also reported for glass microspheres (100 µm and 230 µm diameters) agitated in vibrating bowls [11,12]. In all cases, the stuck particles could be separated by introducing $O_2$ into the sputter chamber, which leads to oxide layers on the already deposited material [11,12].

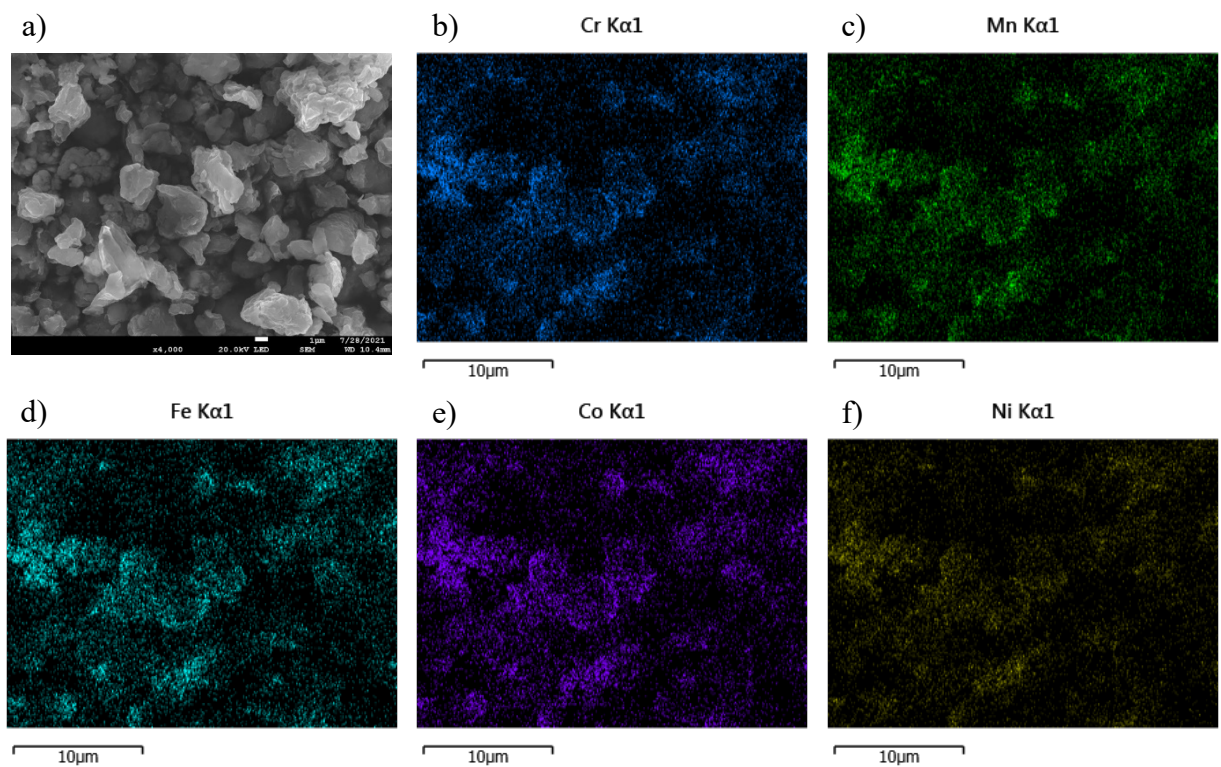

**Fig. 3:** (a) SEM image of exemplary Al particles (5 µm mean size) coated with a CrMnFeCoNi high entropy alloy thin film and the corresponding EDX element distribution maps for (b) Cr, (c) Mn, (d) Fe, (e) Co and (f) Ni.

During the second deposition process, the oscillation plate agitated the spherically shaped polystyrene and Fe alloy particles as well as the cubic NaCl particles to jump up and down. This resulted in an even distribution of the target elements Cr, Mn, Fe, Co and Ni on the particles, as shown in Figs. 4, 5 and 6. The even appearances of the surfaces in the SEM images (Figs. 4a, 5a, 6a) indicate continuous CrMnFeCoNi coatings on the particles. This agrees with the optical image shown in Fig. 7. The average chemical compositions of the films on the polystyrene and the NaCl particles are $Cr_{18.1}Mn_{17.1}Fe_{20.5}Co_{20.9}Ni_{23.4}$ and $Cr_{18.1}Mn_{17.5}Fe_{20.4}Co_{21.4}Ni_{22.6}$, respectively. The compositional variations within the measured particles are very small and the corrected standard sample deviations for the elemental contents are < 0.5 at. %. Considering the approximate EDX accuracy of 1 at. %, it can be concluded that the films on the polystyrene and the NaCl particles have very similar compositions. Compared to the $Cr_{20.6}Mn_{20.4}Fe_{20.1}Co_{19.8}Ni_{19.1}$ sputter target, the films on the polystyrene and NaCl particles contain 2.5 at. % less Cr, 2.9 and 3.3 at. % less Mn, 0.3 and 0.4 at. % more Fe, 1.1 and 1.7 at. % more Co, and 3.3 and 4.5 at. % more Ni, respectively. These differences can be explained by different sputter rates for the different target elements. The composition of the film on the Fe alloy particles was not determined because their Fe content confounds the result.

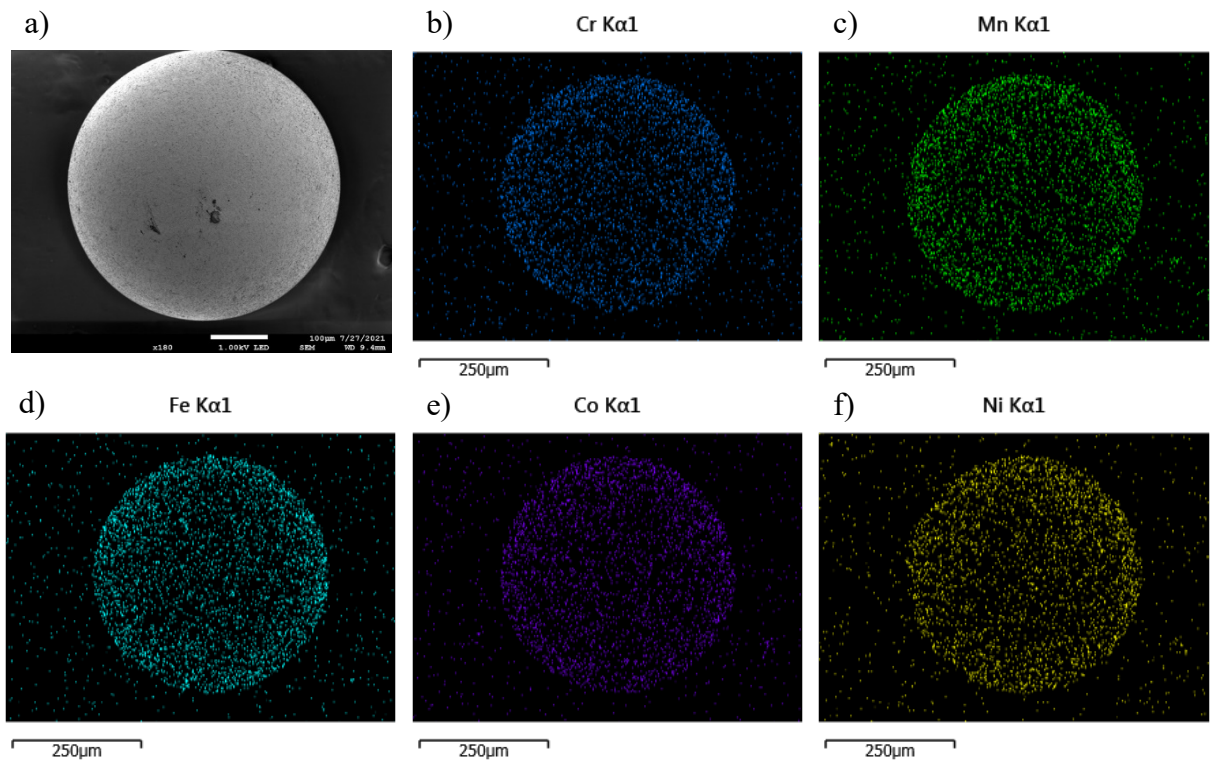

**Fig. 4:** (a) SEM image of an exemplary polystyrene particle coated with a CrMnFeCoNi high entropy alloy thin film and the corresponding EDX element distribution maps for (b) Cr, (c) Mn, (d) Fe, (e) Co and (f) Ni.

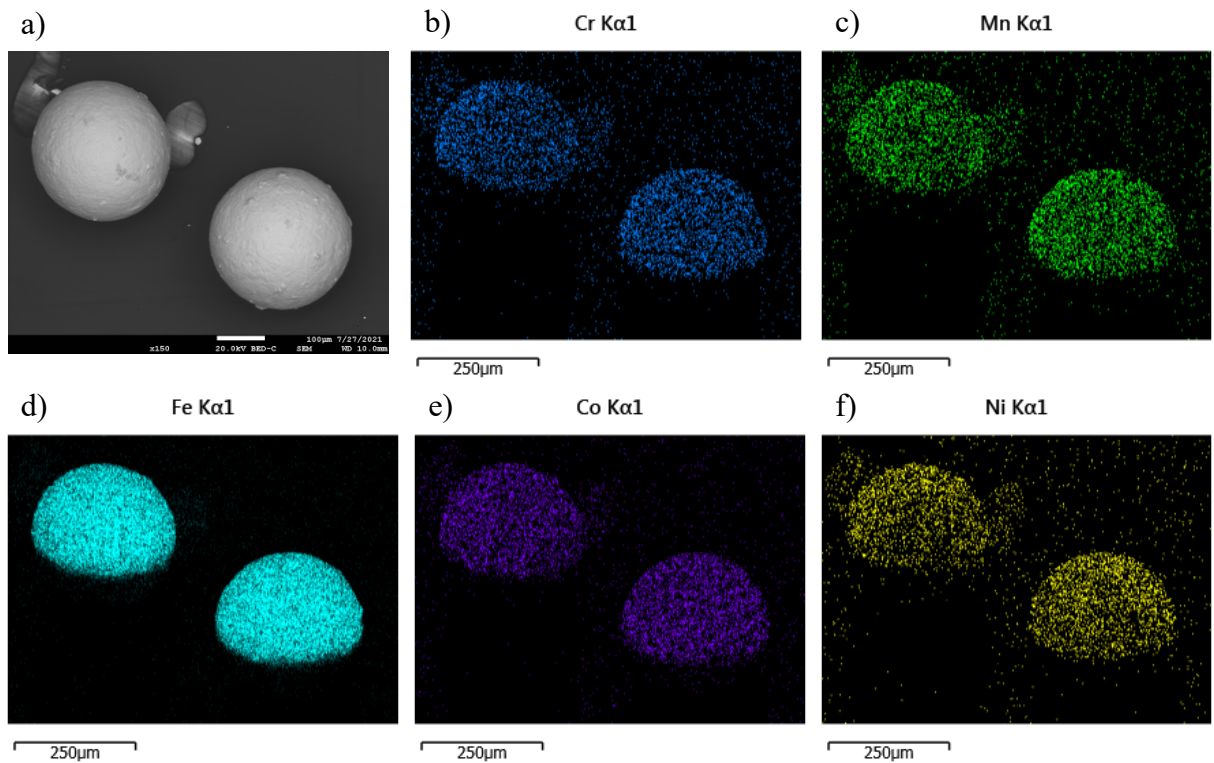

**Fig. 5:** (a) SEM image of exemplary Fe alloy particles coated with a CrMnFeCoNi high entropy alloy thin film and the corresponding EDX element distribution maps for (b) Cr, (c) Mn, (d) Fe, (e) Co and (f) Ni.

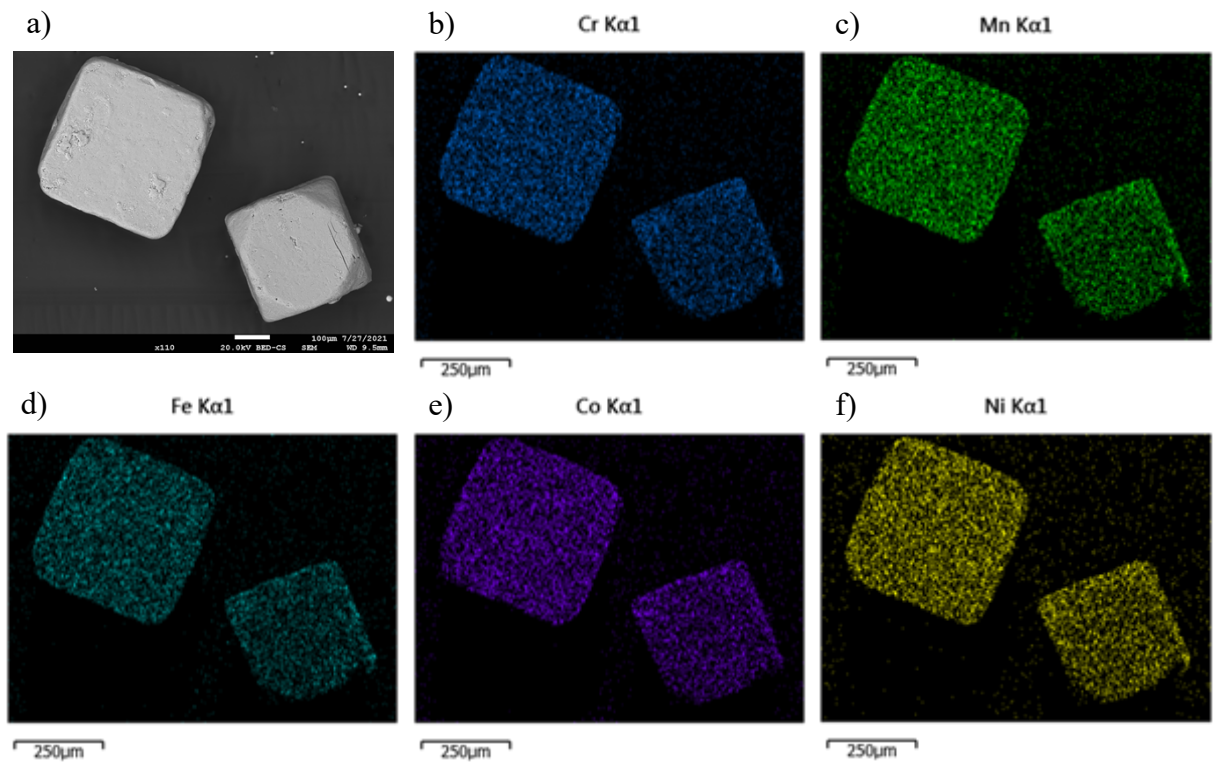

**Fig. 6:** (a) SEM image of exemplary NaCl particles coated with a CrMnFeCoNi high entropy alloy thin film and the corresponding EDX element distribution maps for (b) Cr, (c) Mn, (d) Fe, (e) Co and (f) Ni.

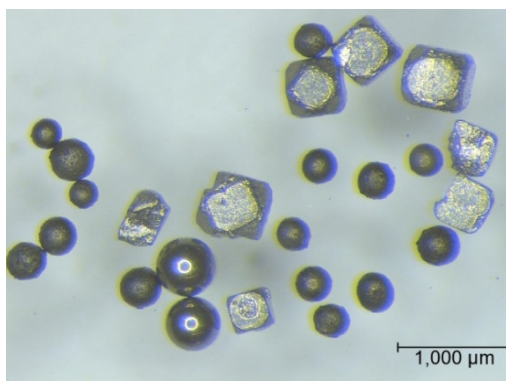

**Fig. 7:** Optical image (Leica M125 C optical stereomicroscope) of exemplary polystyrene, Fe alloy and NaCl particles, that were simultaneously coated with a CrMnFeCoNi high entropy alloy thin film during the second deposition process.

During the third process (co-deposition), the oscillation plate caused all nine batches of Al particles to jump up and down in similar manners. This led to an even distribution of the target elements Cr, Mn, Fe, Co and Ni on the particles, as shown for exemplary particles of batch 9 in Fig. 8. Fig. 9b shows that the co-deposition resulted in a thin film with an average composition of $Cr_{9.1\,...\,17.4}Mn_{9.1\,...\,18.1}Fe_{9.7\,...\,21.2}Co_{10.3\,...\,22.0}Ni_{33.2\,...\,58.1}$, depending on the particle batch. The composition gradient corresponds to the spatial orientation of the particle batches relative to the sputter targets. For instance, the highest and lowest Ni contents were measured for the film on the particles of batches 1 and 9, which were the batches located the closest and the furthest to the Ni target, respectively. Similarly, the highest and lowest Cr and Mn contents were measured for batches 7 and 3, respectively. The highest and lowest Fe and Co contents were measured for batches 9 and 1, respectively. The batches between the ones with maximum and minimum elemental film contents have intermediate compositions. The usage of equiatomic CrMn and FeCo compound targets is reflected by the similar contents of Cr and Mn, and Fe and Co in each batch. The film composition also varies within the batches, since no two particles have coatings of exactly the same composition. Shadow effects caused by the particles of the same batch and the random particle locations within the pits of the oscillation plate are believed to be the reason for that. The average values of the corrected sample standard deviations of the Cr, Mn, Fe, Co and Ni contents in the coatings across the nine particle batches are 3.7 at. %, 3.4 at. %, 3.7 at. %, 4.3 at. % and 9.0 at. %, respectively.

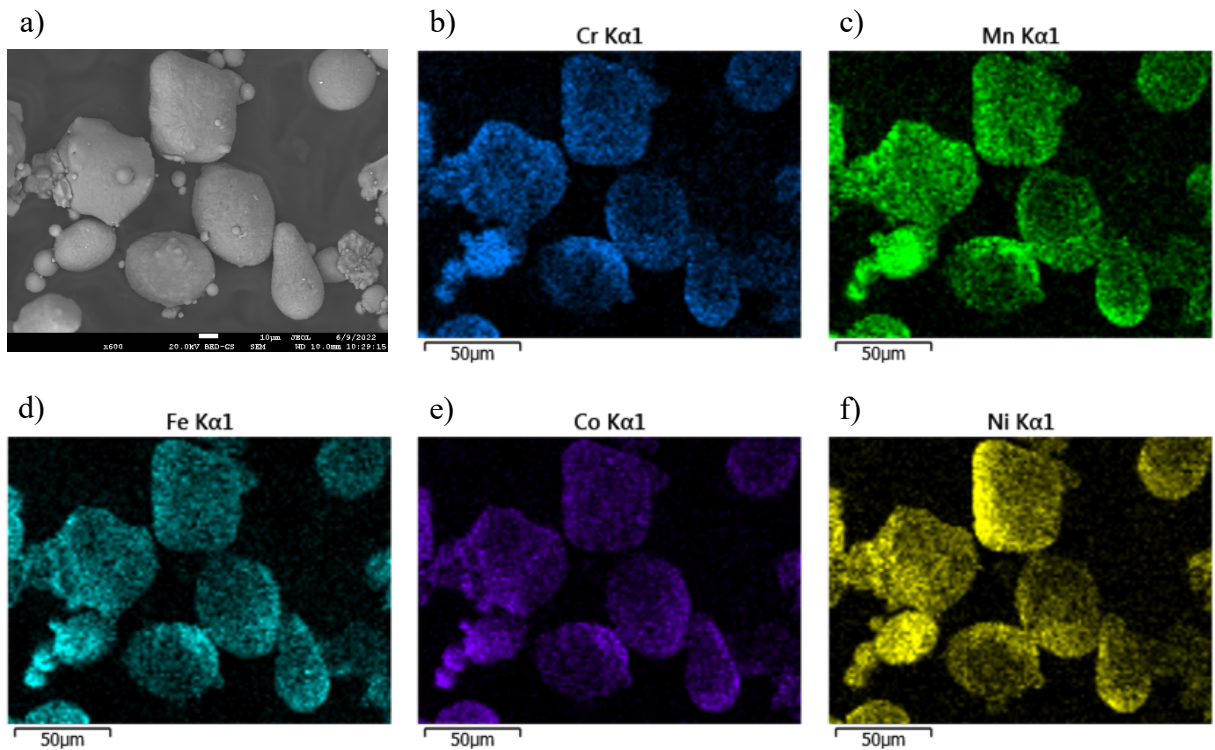

**Fig. 8:** (a) SEM image of exemplary Al particles (25 µm mean size) of batch 9 of the third deposition process, that were coated with a CrMnFeCoNi high entropy alloy thin film, and the corresponding EDX element distribution maps for (b) Cr, (c) Mn, (d) Fe, (e) Co and (f) Ni.

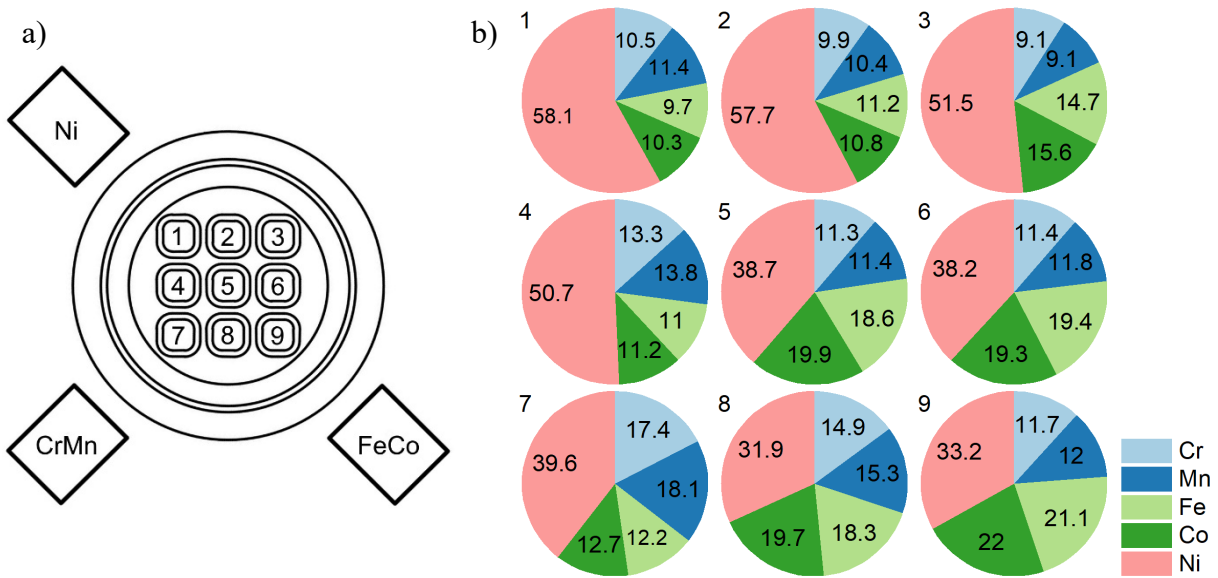

**Fig. 9:** (a) Schematic illustration of the sputter set up of the third process (co-deposition) including the positions of the sputter targets and the numbers one to nine that indicate the batches of Al particles. (b) Visualization of the average chemical compositions (at. %) measured by EDX of the CrMnFeCoNi thin film on ten particles of each particle batch.

## 5. Conclusion and outlook

The presented sputter method was demonstrated to be a promising approach to rapidly produce samples of coated particles with different combinations of core and coating materials. Different kinds of particles were coated with a CrMnFeCoNi thin film, and Al particles were coated with a composition gradient CrMnFeCoNi thin film. Due to the nature of particles of being separate individuals, a composition gradient film on particles inherently has discrete compositions (one composition for each particle coating), whereas films on flat continuous substrates can have continuous composition gradients. To increase the efficiency of the presented sputter method, a scaled-up version of the oscillation plate to accommodate a larger number of particle samples can easily be designed and adapted to any sputter chamber. To increase the efficiency of the sample characterization, high-throughput methods suitable for the fabricated particle samples are recommended to be developed. For the homogeneity of the films, thorough agitation of the particles is necessary. As observed for the agglomerating 5 μm Al particles, insufficient agitation leads to uneven coatings. The oscillation frequency and the travel range are important parameters for the particle agitation, which are controlled by the electrical driving signal. However, to homogeneously coat particles that tend to agglomerate, the method has to be modified. For example, agglomerates may be separated by introducing $O_2$ into the sputter chamber or by using an electrical driving signal with a larger $V_{P-P}$ to cause the actuator to oscillate with a larger travel range.

**Declaration of competing interest**

The authors declare that they have no known competing financial interests or personal relationships that could have appeared to influence the work reported in this paper.

**Acknowledgements**

This research did not receive any specific grant from funding agencies in the public, commercial, or not-for-profit sectors. The Center for Interface Dominated High-Performance Materials (ZGH, Ruhr University Bochum) is acknowledged for the access to the SEM and the optical microscope. We thank Prof. Dr. Guillaume Laplanche (Institute for Materials, Ruhr University Bochum) for providing the CrMnFeCoNi high entropy alloy sputter target.